\documentclass[prl,aps,showpacs,
twocolumn,
]{revtex4-1}

\usepackage{graphicx}
\usepackage{amsmath}
\usepackage{amssymb}
\usepackage{hyperref} 

\newcommand{\be}{\begin{equation}}
\newcommand{\ee}{\end{equation}}
\newcommand{\bea}{\begin{eqnarray}}
\newcommand{\eea}{\end{eqnarray}}

\newcommand{\MKK}{M_{\rm KK}}

\newcommand{\uKK}{u_{\rm KK}}

\newcommand{\Tr}{{\rm Tr}\,}

\newcommand{\D}{G_{D}}

\def\PDG{\cite{Agashe:2014kda}}
\def\BPR{\cite{Brunner:2015oqa}}

\begin{document}

\title{Nonchiral enhancement of scalar glueball decay in the Witten-Sakai-Sugimoto model}

\preprint{TUW-15-XX}

\author{Frederic Br\"unner}
\author{Anton Rebhan}
\affiliation{Institut f\"ur Theoretische Physik, Technische Universit\"at Wien,
        Wiedner Hauptstrasse 8-10, A-1040 Vienna, Austria}

\date{\today}

\begin{abstract}
We estimate the consequences of finite masses of pseudoscalar mesons on the decay rates
of scalar glueballs in the Witten-Sakai-Sugimoto model, a top-down holographic
model of low-energy QCD, by extrapolating from the
calculable vertex of glueball fields and
the $\eta'$ meson which follows from the Witten-Veneziano mechanism
for giving mass to the latter.
Evaluating the effect on the recently calculated
decay rates of glueballs in the Witten-Sakai-Sugimoto model, we find a strong enhancement of the decay of scalar glueballs into
kaons and $\eta$ mesons, in fairly 
close agreement with experimental data on the glueball candidate $f_0(1710)$.
\end{abstract}

\maketitle

The fundamental theory of the strong interactions, quantum chromodynamics (QCD),
which has quarks confined in color-neutral bound states, admits also bound states
whose valence constituents are all gluons, the nonabelian gauge bosons of QCD.
This prediction of additional mesons called gluonia or glueballs dates back to the early 1970's
\cite{Fritzsch:1972jv,*Fritzsch:1975tx,*Jaffe:1975fd} and has been
substantiated by lattice QCD \cite{Morningstar:1999rf,*Chen:2005mg,*Loan:2005ff},
which estimates the mass of the lowest glueball state to be around 1600--1800 MeV.
Experimentally, however, their status remains unclear and controversial
\cite{Bugg:2004xu,*Klempt:2007cp,*Crede:2008vw,*Ochs:2013gi}.
The lowest scalar glueball state has quantum numbers of the vacuum
and can be expected to mix with scalar mesons made from quarks and antiquarks.
To disentangle the contributions, information on decay processes is needed.
Theoretical expectations vary greatly---the lowest glueball state may be 
even so broad that it forms a mere background for the isoscalar
meson spectrum \cite{Minkowski:1998mf}.

QCD in the limit of a large number of colors $N_c$ \cite{'tHooft:1973jz,Lucini:2012gg}, 
which in many cases
turns out to be a remarkably successful approximation to real QCD with $N_c=3$,
predicts a parametric suppression of decay rates of glueballs compared to light quarkonia
by a factor $1/N_c$ as well as a suppression of mixing \footnote{This is not the case in bottom-up holographic models for the Veneziano limit of QCD \cite{Arean:2013tja}, where $N_f/N_c$ is kept fixed as $N_c\to\infty$,  for which Ref.~\cite{Iatrakis:2015rga} recently obtained strong glueball-meson mixing. By contrast, a recent study of a bottom-up hard-wall model obtained only very small mixing \cite{Yamaguchi:2015mha}.}.
If glueballs are indeed narrow and not strongly mixed, one should be able
to identify one of the isoscalar-scalar mesons below 2 GeV
as a predominantly glueball state. In phenomenological studies the experimentally well-established \PDG\
mesons $f_0(1500)$ and $f_0(1710)$  have been identified alternatingly as
possible glueball candidates \cite{Amsler:1995td,*Lee:1999kv,*Close:2001ga,*Amsler:2004ps,*Close:2005vf,*Giacosa:2005zt,*Albaladejo:2008qa,*Mathieu:2008me,*Janowski:2011gt,Janowski:2014ppa,Cheng:2015iaa}.
Both are comparatively narrow states, but their decay patterns are rather different:
$f_0(1500)$ decays primarily into four pions and secondly into two pions, with decays into kaons and $\eta$ mesons
suppressed, whereas $f_0(1710)$ instead decays predominantly into two kaons,
with a ratio \PDG\
$\Gamma(2\pi)/\Gamma(K\bar K)=0.41{+0.11\atop-0.17}$, much lower than
3:4 expected from a flavor-blind glueball.
In the case of $f_0(1500)$ the strong deviation from flavor-blindness
is usually attributed to mixing, while for $f_0(1710)$ it has been suggested that
glueballs couple more strongly to the more massive pseudoscalar mesons,
a mechanism termed ``chiral suppression'' \cite{Sexton:1995kd,Chanowitz:2005du}, which
could make it possible that $f_0(1710)$ is a nearly unmixed glueball as
most recently argued for in \cite{Janowski:2014ppa,Cheng:2015iaa} (see also Ref.~\cite{Gui:2012gx}).

Since lattice QCD results on glueballs in interaction with quarks are still sparse,
in particular concerning decay patterns, it is of interest to employ (top-down) gauge-gravity
duality, a string-theoretic approach to study strongly coupled large-$N_c$ 
gauge theories, 
to obtain new insights from first principles \footnote{Also various bottom-up holographic models
have been used for studying glueball spectra and properties, see e.g.~\cite{BoschiFilho:2002ta,*Colangelo:2007pt,*Forkel:2007ru,Arean:2013tja,Iatrakis:2015rga,Yamaguchi:2015mha}.}. In fact, the spectrum of glueballs
has been one of the first applications of a nonsupersymmetric holographic model derived
by Witten \cite{Witten:1998zw} from type-IIA superstring theory
\cite{hep-th/9805129,*Csaki:1998qr,*Hashimoto:1998if,*Csaki:1999vb,Constable:1999gb,Brower:2000rp}.
The Witten model has subsequently been extended by Sakai und Sugimoto to include
chiral quarks through D8-$\overline{\text{D8}}$ probe branes \cite{Sakai:2004cn,Sakai:2005yt}.
With only one free coupling constant at a (Kaluza-Klein) mass scale $\MKK\sim1$ GeV,
this provides a remarkably successful model for low-energy QCD
with quantitative predictions for vector and axial vector meson spectra and
decay rates that agree with real QCD to within 10-30\% \cite{Rebhan:2014rxa}.

In Ref.~\cite{Hashimoto:2007ze}, the Witten-Sakai-Sugimoto model was used for the first time
to evaluate the decay rate of the lowest glueball state into pions and to compare
with experimental data for the $f_0(1500)$, although the mass of the lowest holographic
glueball mode comes out at 855 MeV.
In Ref.~\cite{Brunner:2015oqa} we have recently revisited this calculation with the result
that the decay width of the lowest mode is much higher than the one obtained in \cite{Hashimoto:2007ze}. Since the lowest mode corresponds to an ``exotic polarization''
\cite{Constable:1999gb} of the gravitational fields of the Witten model, we have
proposed to discard the latter and to instead consider the next-lowest, predominantly dilatonic mode with mass 1487 MeV as corresponding to the glueball in QCD. Despite the closeness of its mass to that of the $f_0(1500)$ meson,
we have found that the decay pattern into two and four pseudoscalar mesons
is not reproduced: the decay rate of $f_0(1500)$ into two pions is underestimated by
about a factor of 2, while the prediction for the dominant decay mode of $f_0(1500)$, 
which is decay into four pions,
is an order of magnitude too small. Extrapolating the mass of the holographic
glueball to that of the glueball candidate $f_0(1710)$ (which is within 16\% of the mass of the dilatonic mode) \footnote{In Ref.~\cite{Brunner:2015oqa} this extrapolation
was made such that the dimensionless ratio of partial width over mass for the decay into 
two massless pseudoscalars is unchanged while taking into account
that the $f_0(1710)$ has a mass above the threshold of two $\rho$ mesons.}, 
we have instead found close
agreement with the decay rate into two pions. Since the Witten-Sakai-Sugimoto model
is chiral, pions, kaons, and $\eta$ mesons are predicted simply in flavor-symmetric ratios 3:4:1,
thus failing to explain the much stronger decays into two kaons and two $\eta$ mesons.

In this Letter we study the possible effect of finite quark masses on the decay
rate of glueballs in the Witten-Sakai-Sugimoto model in order to see whether
a sufficiently strong enhancement of the decay into kaons and also $\eta$ mesons could
result. In Refs.~\cite{0708.2839,*Dhar:2007bz,*Dhar:2008um,*Niarchos:2010ki,Aharony:2008an,*Hashimoto:2008sr,*McNees:2008km}, it has been shown that nonlocal mass terms implementing Gell-Mann-Oakes-Renner relations can be induced
by either worldsheet instantons or a deformation by a bifundamental field
related to the open string tachyon that arises between (parallel) D and anti-D branes.
Since no complete calculation along these lines exists, the additional coupling
of glueballs and pseudoscalar mesons induced by the nonlocal mass term is not known.
However, the additional coupling of glueballs to $\eta'$ mesons
due to the part of its
mass term which arises from the anomalous breaking of U(1)$_A$ flavor symmetry
can be calculated exactly. 
We propose to use that as a simple and also plausible model for how the totality of
nonlocal mass terms for the pseudoscalar meson depend on the scalar glueball fields,
and to thereby
extrapolate the results
for the glueball decay pattern obtained in \cite{Brunner:2015oqa}
to finite pseudoscalar masses.

In the chiral Witten-Sakai-Sugimoto model \cite{Sakai:2004cn,Sakai:2005yt},
the coupling of pseudoscalar Goldstone
bosons as well as vector and axial vector mesons to scalar and tensor glueball
fields are determined by the dependence of the Dirac-Born-Infeld part of the D8 brane action on
metric and dilaton fields,
\be\label{SD8}
S_{\rm D8}^{\rm DBI}=-
T_{\rm D8}
\Tr\int d^9x e^{-\Phi}
\sqrt{-\det\left(\tilde g_{MN}+2\pi\alpha' F_{MN}\right)},
\ee
where $\Phi$ and $\tilde g_{MN}$ are the dilaton and the nine-dimensional induced metric in
the ten-dimensional background given by 
\bea\label{ds2W}
&& ds^2=\left(\frac{u}{R}\right)^{3/2}\left[\eta_{\mu\nu}dx^\mu dx^\nu+f(u)dx_4^2\right]\nonumber\\
&&\quad+\left(\frac{R}{u}\right)^{3/2}\left[\frac{du^2}{f(u)}+u^2 d\Omega_4^2\right], 
\; f(u)=1-\frac{\uKK^3}{u^3},
\\
&&e^\Phi=g_s \left(\frac{u}{R}\right)^{3/4},
\eea
with circle-compactified $x^4\simeq x^4+2\pi/\MKK$ and $\MKK=\frac32 {\uKK^{1/2}}{R^{-3/2}}$.
The stacks of $N_f$ D8 and anti-D8 branes are assumed to be
localized at antipodal points, giving rise to
trivial embeddings $x^4=const.$, which extend from the holographic boundary
at $u=\infty$ to the minimal point $\uKK$ where branes and antibranes connect.
This breaks the chiral group $\mathrm{U}(N_f)_L\times \mathrm{U}(N_f)_R$ down
to its diagonal group,
leading to a nonet (for $N_f=3$) of pseudoscalar Goldstone bosons
described by
\be\label{UPi}
U(x)=\mathrm P\,\exp i\int_{-\infty}^\infty dz A_z(z,x)=e^{i\Pi^a\lambda^a/f_\pi},
\ee 
where $z/\uKK=\sqrt{(u/\uKK)^3-1}$ parametrizes the radial extent of the joined
D8 and anti-D8 branes.
Matching the pion decay constant $f_\pi$ to 92.4 MeV and the mass of the lowest
vector meson mode $A_\mu(z,x)=\rho_\mu(x)\psi_{(1)}(z)$ to the $\rho$ meson mass
$m_\rho\approx 776$ MeV fixes $\MKK=949$ MeV and $\lambda=g_{\rm YM}^2N_c=16.63$
\cite{Sakai:2004cn,Sakai:2005yt}; matching instead $m_\rho/\sqrt\sigma$ with
$\sigma$ the string tension of the model to large-$N_c$ lattice results 
\cite{Rebhan:2014rxa,Brunner:2015oqa}
gives a somewhat lower value of the 't Hooft coupling, $\lambda=12.55$, which
we use with the higher value to give a band of variation for the holographic
predictions.

In Ref.~\cite{Barbon:2004dq,Sakai:2004cn} (see also \cite{Armoni:2004dc}) it was shown that the U(1)$_A$ anomaly
requires to combine the Ramond-Ramond 2-form field strength $F_2$ with the isoscalar
$\eta_0$ that is localized on the D8 branes in a gauge-invariant combination $\tilde F_2$
with bulk action
\be\label{SC1}
S_{C_1}=-\frac1{4\pi(2\pi l_s)^6}\int d^{10}x\sqrt{-g}|\tilde F_2|^2
\ee
with
\be\label{Ftilde2}
\tilde F_2=\frac{6\pi\uKK^3 \MKK^{-1}}{u^4}
\left(\theta+\frac{\sqrt{2N_f}}{f_\pi}\eta_0\right)du \wedge dx^4,
\ee
where $\theta$ is the QCD theta angle and
\be
\eta_0(x)=\frac{f_\pi}{\sqrt{2N_f}}\int dz \,{\rm Tr} A_z(z,x).
\ee
This gives rise to a Witten-Veneziano \cite{Witten:1979vv,Veneziano:1979ec} mass term for $\eta_0$ that
is local with respect to the effective 3+1-dimensional
boundary theory but nonlocal in the bulk, with mass squared
\be
m_{0}^2=\frac{N_f}{27\pi^2 N_c}\lambda^2\MKK^2.
\ee
For $N_f=N_c=3$, $\MKK=949$ MeV, and $\lambda$ varied from 16.63 to 12.55 one finds
$m_{0}=967\ldots730$ MeV.

The other pseudoscalar mesons described by (\ref{UPi}) are massless
in the Witten-Sakai-Sugimoto model. Current quark masses can in principle
be introduced through a deformation by a bulk field $\mathcal T$ in the
bifundamental representation of the chiral symmetry group \cite{0708.2839,*Dhar:2007bz,*Dhar:2008um}
that is related to tachyon condensation 
or alternatively through worlsheet instantons
\cite{Aharony:2008an,*Hashimoto:2008sr,*McNees:2008km}.
Both introduce nonlocal mass terms for the pseudoscalar mesons, which
one may qualitatively write as
\be\label{calT}
\int d^4x \int_{\uKK}^\infty du\,h(u)\,\Tr\left(\mathcal T(u)\,\mathrm P\, e^{-i\int dz A_z(z,x)}+h.c.\right),
\ee
where $h(u)$ includes metric fields.
Choosing appropriate boundary conditions for $\cal T$,
the quark mass matrix arises through
\be
\int_{\uKK}^\infty du\,h(u)\,\mathcal T(u) \propto \mathcal M={\rm diag}(m_u,m_d,m_s),
\ee
thereby realizing a Gell-Mann-Oakes-Renner relation.

Integration over $u$ leads to mass terms for all Goldstone bosons including one for the flavor singlet $\eta_0$
in addition to the Witten-Veneziano mass term. The flavor octet $\eta_8$ and $\eta_0$ can be diagonalized
to mass eigenstates $\eta$ and $\eta'$. 
With $\mathcal M={\rm diag}(m,m,m_s)$, $m=(m_u+m_d)/2$, fixing $m_\pi=140$ MeV and $m_K=497$ MeV, this diagonalization
yields for $\lambda=16.63\ldots12.55$:
\bea
&&m_\eta=518\ldots476\; {\rm MeV},\quad 
m_{\eta'}=1077\ldots894\; {\rm MeV},\quad\\
&&\theta_P=-14.4^\circ\ldots-24.2^\circ,
\label{thetaP}
\eea
with $\theta_P$ the octet-singlet mixing angle,
which shows that the above holographic result for $m_{0}$ is in the right ballpark \footnote{This setup for the masses is 
incapable of actually matching the experimental
mass ratio $m_\eta/m_{\eta'}$ for any value of $m_u,m_d,m_s,m_{0}$
\cite{Georgi:1993jn}, but $O(p^4)$ corrections involving $\mathcal M$
can fill the deficit \cite{Gerard:2004gx,Mathieu:2010ss}.
The mixing angle (\ref{thetaP}) has been shown to receive only small corrections of this kind \cite{Gerard:2004gx}.
Values around $-14^\circ$ appear to be favored by data on light meson decays \cite{Ambrosino:2009sc,Pham:2015ina}, 
while radiative charmonium decay
instead points to $\theta_P\approx -21^\circ$ \cite{Gerard:2004gx,Gerard:2013gya}.}.

In order to determine how mass terms affect the coupling of glueballs and mesons
worked out in Ref.~\cite{Brunner:2015oqa}, 
we would need to know
the dependence on dilaton and metric fields of $h(u)$ as well as the profile of
the bifundamental field $\mathcal T(u)$.
Absent those, we turn to the fully known nonlocal mass term
produced by (\ref{SC1}) and (\ref{Ftilde2}) for $\eta_0$.
Inserting the mode expansion of glueball fields $G_D$ and $G_E$
defined in Ref.~\cite{Brunner:2015oqa}, we find
\be
S^{\rm eff.}_{\eta_0}=-\frac12 \int d^4x\, m_{0}^2 \,\eta_0^2 \left( 1-3 d_0 G_D+5
\breve c_0 G_E\right)+\ldots
\ee
with
\bea\label{d0}
d_0&=&3\uKK^3\int_{\uKK}^\infty H_D(u) u^{-4} du
\approx \frac{17.915}{\lambda^{1/2}N_c\MKK},\\
\label{cbreve0}
\breve c_0&=&\frac34 \uKK^3\int_{\uKK}^\infty H_E(u) u^{-4} du
\approx  \frac{15.829}{\lambda^{1/2}N_c\MKK},
\eea
where the latter is given for completeness only, since we
are going to discard the ``exotic'' mode $G_E$ given the results in \BPR.
Here $H_{D,E}(u)$ are the radial profile functions of the glueball modes, normalized
such as to give a canonical kinetic term for $G_{D,E}(x)$.

Given the similarity to how a nonlocal mass term is generated through worldsheet
instantons, this result seems to be a reasonable first guess of how nonlocal mass terms
couple in general. 
As far as a bifundamental field $\mathcal T$ associated with tachyon condensation is concerned, a plausible guess would be that the metric dependence
derives from the integration measure of D8 branes, $d^9x\, e^{-\Phi}\sqrt{-\tilde g}.$
For the predominantly dilatonic glueball field \footnote{For the ``exotic'' scalar glueball field,
the dependence is different. In the notation of \BPR, it leads to a factor $[1+(H_E-\bar H_E)G_E]$.}, this turns out to have exactly
the same dependence on terms linear in $G_D(x)H_D(u)$ as follows from (\ref{SC1}) and (\ref{Ftilde2}),
namely a factor $(1-3G_D(x)H_D(u))$.
In order to calculate the coupling constant analogous to $d_0$ in (\ref{d0}),
one would need to know the holographic profile of $\mathcal T$, of which we only
know that it will be concentrated around $u=\uKK$. As a simplistic guess
one could try a function that mimics the profile of the term $A_z \partial_\mu^2 A_z$
in the D-brane action when $A_z$ equals the zero mode describing the Goldstone bosons. 
This would simply determine the analog of $d_0$ to be equal
to the coupling $d_1$ appearing in the chiral $G_D\pi\pi$ term \BPR,
\be\label{LGDchiral}
\mathcal L_{\D\pi\pi}^{\rm chiral}= \frac{1}{2}d_1 \Tr (\partial_\mu\pi\partial_\nu\pi)
\left(\eta^{\mu\nu}-\frac{\partial^\mu \partial^\nu}{M^2}\right)\D,
\ee
where $d_1\approx 17.226 \lambda^{-1/2}N_c^{-1}\MKK^{-1}$.
This differs from $d_0$ by a mere 4\%.

We shall therefore continue with the working hypothesis
that the overall coupling of the glueball field to the mass term for the pseudoscalar mesons
is universal.
This essentially assumes
that the mixing of singlet and octet mesons is invariant under
a holographic renormalization group evolution, 
which in particular implies the absence of a direct coupling $G\eta\eta'$.

With this assumption, i.e.\ adding
\be\label{mPPG}
-\frac12\sum_i m_i^2 P_i^2(1-3d_0 G_D)
\ee
with $P_i$ the mass eigenstates of the pseudoscalar mesons
to (\ref{LGDchiral}), we obtain the following
modification factor for the decay rate to two pseudoscalar mesons of mass $m_P$:
\be\label{enhancement}
\left(1-4\frac{m_P^2}{M^2}\right)^{1/2}\left(1+\alpha \frac{m_P^2}{M^2}\right)^2
\ee
with
\be\label{alphaD}
\alpha=4(3 d_0/d_1 - 1)= 8.480\ldots \text{ for $G_D$}.
\ee
An analogous calculation for the ``exotic'' scalar glueball using
the results of \BPR\ for the chiral contributions gives
\be\label{alphaE}
\alpha=4(5\breve c_0-\breve c_1)/(c_1+2\breve c_1)=2.630\ldots \text{ for $G_E$}.
\ee
[Note that to leading order the dependence on $\lambda$ and $\MKK$ drops out in (\ref{alphaD}) and (\ref{alphaE}).]
In (\ref{enhancement}) the first factor represents a simple kinematical suppression
which is overcome by the coupling of the glueball field to the mass term of
the pseudoscalar fields. A similar result, but with $\alpha=1$ was
obtained in Ref.~\cite{Ellis:1984jv} for a simple effective field theory
where the scalar glueball field is identified with the dilaton of QCD (a scalar
field with a potential matched to the QCD trace anomaly).
With $\alpha=1$ the nonchiral enhancement is cancelled by the kinematical suppression
to order $m_P^2/M^2$, thus restoring approximate flavor symmetry, while for
larger values $m_P$ the net effect is a (slight) reduction of the decay rate.

In Table \ref{tabratios} we compare the deviations from flavor symmetry as they
are reported by the Particle Data Group \PDG\ with the modification resulting
from (\ref{enhancement}) and (\ref{alphaD}).
Remarkably, the experimental ratio $\Gamma(\pi\pi)/\Gamma(K\bar K)$ is
reproduced within the experimental error bar, whereas the
prediction for $\Gamma(\eta\eta)/\Gamma(K\bar K)$ remains within 1.33 standard deviations.

In Table \ref{tabrates} we compare our complete set of predictions for the
decay rates for a scalar glueball with mass corresponding to either
1505 MeV ($f_0(1500)$) or 1722 Mev ($f_0(1710)$), with and without inclusion of
masses for the pseudoscalar mesons, to experiment.
In the case of $f_0(1500)$, the (experimentally well-known) decay pattern is
neither matched qualitatively nor quantitatively \BPR. Inclusion of the pseudoscalar masses
helps for the total width but modifies the decay pattern adversely.
For $f_0(1710)$, branching ratios are less accurately known.
The prediction for the total width is increased slightly above the experimental
value when masses are included, but the decay pattern into two pseudoscalar mesons
is improved markedly, as we have already shown in Table \ref{tabratios}.
In fact, the experimental results quoted for the partial widths should
be considered as upper values as they ignore the possibility of decay into
four or more pions, since at least decay of $f_0(1710)$ into two $\omega$ mesons and
further to six pions has been seen \PDG\
in $J/\psi\to\gamma f_0(1710)$ (in fact at the level of 75\% of the rate into two pions
\PDG, so that the holographic prediction may not be very far off) \footnote{This
interpretation of the data published in \cite{Ablikim:2006ca} ignores
the possibility that these decays could be attributed to a separate isoscalar 
such as the $f_0(1790)$ proposed in \cite{Ablikim:2004wn}.}
The only remaining major mismatch between existing experimental data for $f_0(1710)$ \cite{Close:2015rza} and
the prediction of the Witten-Sakai-Sugimoto model thus appears to be
the rather high rate for decay into four pions, which is predicted by the latter to proceed through
$2\rho$ and $\rho\pi\pi$ at the level of
about twice the rate into two $\omega$ mesons \BPR, and this prediction
is not modified by the introduction of quark masses at the level of our approximation.

To summarize, by extrapolating the exactly calculable coupling of scalar glueballs
to the mass term of the isosinglet pseudoscalar meson in
the originally chiral Witten-Sakai-Sugimoto model, we found a significantly enhanced
decay of scalar glueballs into kaons and $\eta$ mesons compared to flavor-symmetric ratios.
This is in line with the previously proposed mechanism of ``chiral suppression''
of scalar glueball decay which has been posited as explanation how the isoscalar
meson $f_0(1710)$ with its preferred decay into two kaons 
could be predominantly gluonic rather than an $s\bar s$ state \cite{Sexton:1995kd,Chanowitz:2005du}.
From this we conclude that
the top-down holographic Witten-Sakai-Sugimoto model may well be consistent with
a glueball interpretation of $f_0(1710)$ while disfavoring the other popular glueball candidate $f_0(1500)$.
In this case, the successful reproduction of the branching ratios given in Table I is
correlated to a sufficiently small rate for the decay $G\to\eta\eta'$, for which
only upper limits exist so far \footnote{Relaxing our assumption of a universal coupling (\ref{mPPG})
leads to a nonvanishing coupling $G\eta\eta'$ and a modification of the result (\ref{enhancement}).
A systematic quantitative analysis of this correlation
will be presented in a forthcoming publication.}. 
Moreover, the Witten-Sakai-Sugimoto model
predicts significant partial widths for the decay of $f_0(1710)$ into four and six pions,
of which so far only the latter have been confirmed experimentally according to \PDG.

\begin{table}[t]
\begin{tabular}{llcc}
\toprule
$f_0(1710)$ & exp.(PDG) & WSS massive \\
\colrule
$\frac43 \cdot \Gamma(\pi\pi)/\Gamma(K\bar K)$ &  0.55$+0.15\atop-0.23$  &  0.463 \\
$4 \cdot\Gamma(\eta\eta)/\Gamma(K\bar K)$ &  1.92$\pm0.60$  &  1.12 \\
\botrule
\end{tabular}
\caption{Flavor-asymmetric deviation of branching ratios of glueball candidate $f_0(1710)$ 
compared to the nonchiral enhancement in the Witten-Sakai-Sugimoto model 
resulting from (\ref{enhancement}) and (\ref{alphaD})
with $m_P=m_{\pi,K,\eta}=\{140,497,548\}$~MeV.}
\label{tabratios}
\end{table}

\begin{table*}
\begin{tabular}{lrrr}
\toprule
decay &  $\Gamma/M$ {(exp. \protect\PDG)}  & (WSS chiral \protect\cite{Brunner:2015oqa}) & (WSS massive)\\
\colrule
$f_0(1500)$ (total) & 0.072(5)  & 0.027\ldots0.037 & 0.057\ldots0.077 \\
$f_0(1500)\to4\pi$ & 0.036(3)  &  0.003\ldots 0.005 &  0.003\ldots 0.005 \\
$f_0(1500)\to2\pi$ & 0.025(2)  & 0.009\ldots0.012 & 0.010\ldots0.014\\
$f_0(1500)\to 2K$ & 0.006(1)  & 0.012\ldots0.016 & 0.034\ldots0.045\\
$f_0(1500)\to 2\eta$ & 0.004(1)  & 0.003\ldots0.004 & 0.010\ldots0.013\\
\colrule
$f_0(1710)$ (total) & 0.078(4) & 0.059\ldots0.076 & 0.083\ldots0.106 \\
$f_0(1710)\to 2K$ & * $\left\{ 0.041(2) \atop 0.047(17) \right.$ & 0.012\ldots0.016 & 0.029\ldots0.038 \\[8pt]
$f_0(1710)\to 2\eta$ & * $\left\{0.020(10) \atop 0.022(11) \right.$ & 0.003\ldots0.004 & 0.009\ldots0.011 \\[8pt]
$f_0(1710)\to2\pi$ & * $\left\{0.017(4) \atop 0.009(2) \right.$ & 0.009\ldots0.012 & 0.010\ldots0.013 \\[8pt]
$f_0(1710)\to2\rho,\rho\pi\pi\to4\pi$ & ? & 0.024\ldots 0.030 & 0.024\ldots 0.030 \\ 
$f_0(1710)\to2\omega\to6\pi$ & seen & 0.011\ldots 0.014 & 0.011\ldots 0.014 \\ 
\botrule
\end{tabular}
\caption{Experimental data for the decay pattern of the glueball candidates $f_0(1500)$ and
$f_0(1710)$ from Ref.~\PDG, except for those marked by a star, which
are from Ref.~\cite{1208.0204} where the total width of $f_0(1710)$ was split under the assumption of
a negligible branching ratio to four or more pions,
using data from BES \cite{hep-ex/0603048} (upper entry) and WA102 \cite{hep-ex/9907055} (lower entry),
respectively, compared to the prediction obtained in Ref.~\BPR\ from mode $G_D$ in the chiral Witten-Sakai-Sugimoto model, and finally to the extrapolation of the massive case proposed in this Letter.}
\label{tabrates}
\end{table*}

\begin{acknowledgments}
We thank Denis Parganlija and Timm Wrase for many useful discussions,
and Jean-Marc G\'erard for correspondence.
A.R.\ also thanks the Helsinki Institute of Physics, where this
work was finalized, for hospitality
as well as stimulating discussions.
This work was supported by the Austrian Science
Fund FWF, project no. P26366, and the FWF doctoral program
Particles \& Interactions, project no. W1252.
\end{acknowledgments}

\raggedright
\bibliographystyle{apsrev4-1}
\bibliography{glueballdecay}

\end{document}